# *Solving partial differential equations with waveguide-based metatronic networks*

*Ross Glyn MacDonald[1,2], Alex Yakovlev[2] and Victor Pacheco-Peña[1*]*

[1]*School of Mathematics, Statistics and Physics, Newcastle University, Newcastle Upon Tyne, NE1 7RU, United Kingdom*
[2]*School of Engineering, Newcastle University, Newcastle Upon Tyne, NE1 7RU, United Kingdom*
*email: victor.pacheco-pena@newcastle.ac.uk*

**Abstract**

Photonic computing has recently become an interesting paradigm for high-speed calculation of computing processes using light-matter interactions. Here, we propose and study an electromagnetic wave-based structure with the ability to calculate the solution of partial differential equations in the form of the Helmholtz wave equation, $\nabla^2 f(x,y) + k^2 f(x,y) = 0$, with $k$ as the wavenumber. To do this, we make use of a network of interconnected waveguides filled with dielectric inserts. In so doing, it is shown how the proposed network can mimic the response of a network of T-circuit elements formed by two series and a parallel impedance, i.e., the waveguide network effectively behaves as a metatronic network. An in-depth theoretical analysis of the proposed metatronic structure is presented showing how the governing equation for the currents and impedances of the metatronic network resembles that of the finite difference representation of the Helmholtz wave equation. Different studies are then discussed including the solution of partial differential equations for Dirichlet and open boundary value problems, demonstrating how the proposed metatronic-based structure has the ability to calculate their solutions.

## Introduction

Partial differential equations (PDEs) are fundamental in every area of mathematics, physics, and engineering. They are used to describe many physical phenomena ranging from heat transfer[1] to fluid flow[2] and electromagnetic (EM) wave propagation[3], among others. However, aside from some simplified cases (such as EM planewave propagation[3] or the eigenfunctions/values of a two-dimensional harmonic oscillator[4]) closed form solutions of arbitrary PDEs may be challenging or even impossible to find. Instead, the solution to these equations is usually calculated using numerical techniques such as the finite difference[5] or Finite Element Method (FEM)[6]. Many efforts have been focused on the optimization of these numerical algorithms, such as mesh refinement[7] and parallelization[8], to reduce calculation times and power consumption. Even so, the inherent size and iterative nature[9,10] of these calculations make it a computationally intensive task[11], via conventional computing systems.

In addition to numerical techniques to solve PDEs using computer algorithms (software), recently, computing with matter, i.e., EM wave-based analogue computing, has been suggested as an alternative paradigm for high speed computing[12]. As solutions to computing tasks are calculated by controlling the propagation of EM signals through a material-based analogue processor[13], fast solutions of mathematical operations can be achieved as computing processes are carried out at the speed of light within the materials. In recent works, EM wave-based analogue processors have been demonstrated to perform operations such as integration[14,15], differentiation[13,16,17], convolution[15,18], matrix multiplication[19] and ordinary differential equation[20] (ODE) solving. This has been done by carefully designed structures such that they are able to apply mathematical operators directly onto the wavefront of an incident signal in either space[21–23] or time[13,24,25]. Various design approaches have been explored in order to realize these devices including Bragg gratings[24], diffractive networks[26], Mach-Zehnder interferometers[25], and the application of metamaterials[12,22,27] which enable the arbitrary control over EM wave propagation in both space and time[28–33].

When considering solving PDEs without using software, one can use networks of lumped circuit elements (i.e., resistors, inductors, and capacitors) arranged in a grid. In doing so, this grid can indeed emulate the mesh elements used in the FEM to approximate solutions to various PDEs[34,35]. Interestingly, it has been recently demonstrated that it is also possible to apply the same principle to photonic computing[36–38] systems by exploiting the splitting and superposition of EM signals within an engineered network of dielectric waveguide junctions, called Photonic Kirchhoff nodes[36]. At these nodes, a combination of photonic structures (ring resonators[39,40] and X-junctions[41,42]) can be exploited to emulate the performance of traditional lumped elements in circuit theory. In so doing, it is possible to achieve the splitting of EM waves required to correctly calculate the solution of a PDE such as the Poisson equation, $\nabla^2 f(x,y) = 0$[34–36] (where $\nabla^2 = \partial^2/\partial x^2 + \partial^2/\partial y^2$ is the two-dimensional Laplace operator, $x, y$ are independent parameters and $f$ is the function to be solved). Another interesting approach is to exploit metatronic elements[43] which consist of sub-wavelength metal or dielectric inserts which can emulate the performance of lumped circuit elements within the EM domain. Similar to electronics where circuit elements control the flow of current, when considering EM waves, metatronic elements manipulate the flow of displacement current[44]. Remarkably, this allows much of the knowledge gained from the field of electronics to be effectively transferred to smaller scales using EM waves. This strategy has recently been exploited to aid in the design of filters[45,46], antennas[47,48], metasurfaces[49] and analogue processors[15]. Indeed, metatronic elements using an epsilon-near-zero (ENZ) medium[50–52] have also been exploited to solve PDEs[53]. It has been recently shown how ITO (Indium Tin Oxide) working in the ENZ regime performs an analogous function to a wire in electronics[43] given that the wavelength inside is, effectively, almost infinite. Hence, by immersing metatronic elements (dielectric or metallic inserts) within such ENZ material (host medium), one can emulate the performance of lumped elements within a circuit.

Inspired by the interesting features of metatronic elements emulating circuits with EM waves, here we propose and study an EM wave-based structure for analogue computing with the ability to produce solutions of the Helmholtz wave equation of the form $\nabla^2 f(x,y) + k^2 f(x,y) = 0$, where $k$ is the wavenumber of the PDE to be solved. Different to previous works[53], we make use of a network of parallel plate waveguides without the need of implementing an ENZ host medium. This is achieved by considering that the waveguides are filled with air and loaded with carefully designed dielectric slabs. In so doing, the whole structure acts as metatronic elements which emulate both series and parallel lumped circuit elements. The metatronic elements consist of thin dielectric films separated via a distance of $\lambda_0/4$ (where $\lambda_0$ is the operating wavelength of the network in free-space) (see Fig. 1b). As it is known, and as it will be further explained below, this enables an impedance transformation, allowing for parallel metatronic elements to act as series components[45,46]. A full physical and mathematical analysis of the designed structure is presented, demonstrating how, by controlling the effective impedance values of the metatronic elements, it is possible to tailor the response of the system allowing us to calculate the required PDE solution. To validate the proposed structure working at microwave frequencies, full-wave numerical simulations (from now on referred to numerical solutions) are carried out providing solutions of a range of Dirichlet boundary value problems. Examples include radiation from a dipole, standing waves in a cavity, and focusing/lensing. To fully compare our results, all the numerical solutions are compared with analytical (calculated using computer software-based solution on a traditional FEM technique or using the Huygens-Fresnel principle[54,55]) and theoretical (by solving the equations that govern a perfect metatronic grid) results. We envision that these devices may see application as a computational accelerator in order to produce a fast approximation of a solution to a given PDE. Furthermore, this research may enable the design of additional processors capable of solving higher order PDEs based on the finite difference method.

**Results**

**Solving PDE equations via circuit models: theory**

In previous works, it has been shown how the solution of PDEs (such as the Poisson equation $\nabla^2 g = 0$, where $g$ is the function to be solved for) can be computed by exploiting a network of lumped circuit elements[34,36,38,53,56] connected together in a grid-like lattice. In those works, [34,53] the structure consisted of a unit cell made of periodic junctions of lumped elements (see also Supplementary materials). Each junction can be formed by four lumped elements having an impedance $Z_L$, which are, each of them, connected to adjacent junctions, in this way forming a grid. The governing equation for the voltage distribution of such lumped element-based network can be found by considering the division of current at the junctions according to Kirchhoff's current law, as follows (see Supplementary materials for further details):

$$\frac{1}{Z_L}(V_1 + V_2 + V_3 + V_4 - 4V_0) = 0 \tag{1}$$

where $V_0$ is the voltage at a junction and $V_1$, $V_2$, $V_3$ and $V_4$ are the voltages at the adjacent junctions (top, right, bottom, and left junctions, respectively). To better understand how such lumped circuit network can be exploited for PDE solving, one can compare Eq. 1 with the well-known finite difference approximation of the two dimensional (2D) Laplacian[5,57], which is defined as follows:

$$\nabla^2 g \approx \frac{1}{h^2}[g(x+h,y) + g(x-h,y) + g(x,y+h) + g(x,y-h) - 4g(x,y)] = 0 \qquad (2)$$

where $x$ and $y$ are the independent variables of the function $g$ (such as spatial coordinates) and $h$ is the separation between the sampling points of $g$. It can be seen how Eq. 1 and Eq. 2 are indeed similar provided that the impedance, $Z_L$, of the lumped element for the network is selected such that $Z_L = h^2$. This means that the network of lumped circuit elements can be considered as a simulation space in which the junctions between elements are the sampling points while the impedances of the elements between them control the scaling of the system, i.e., how far apart each point is from one another within the simulation space. With this in mind, by substituting $Z_L = h^2$ into Eq. 1, the voltage distribution of the impedance network, sampled at the junctions of lumped elements, will satisfy Eq. 2. Thus, it may be exploited to calculate a solution of physical problems that are defined via Poisson equation ($\nabla^2 g = 0$), subject to the boundary conditions placed upon the system at the outermost junctions of the grid. The particular example of the PDE $\nabla^2 g = 0$ has multiple uses in modeling problems such as heat transfer[54] or electro/magnetostatics[3,57], as demonstrated in[53]. Another PDE with interesting applications is the Helmholtz equation which has the form $\nabla^2 g + k^2 g = 0$, with $k^2$ as the coefficient of the zeroth-order term of the PDE. This equation can describe EM systems in steady state[3] (where $k$ plays the role of wavenumber). Conceptually, the above method for solving PDEs (described in Eq. 1) may be extended to solve the Helmholtz wave equation by altering the equivalent circuit of the network to include the additional $k^2 g$ term. Based on this, the finite difference representation of the Helmholtz wave equation can be written by expanding the $\nabla^2 g$ term using Eq. 2, as follows:

$$\frac{1}{h^2}[g(x+h,y) + g(x-h,y) + g(x,y+h) + g(x,y-h) - 4g(x,y)] + k^2 g = 0 \qquad (3)$$

Now, what would be an equivalent circuit model that could emulate Eq. 3? To answer this, in this work we propose the circuit shown in Fig. 1a. It consists of a square lattice of T-circuits connected in a series configuration to form a junction. As observed, and in line with other works using lumped elements[34,53,56], the proposed circuit is periodic. However, the junctions are now changed from four interconnected lumped elements to four interconnected T-circuits. As it will be discussed below, the impedances forming the T-circuits can be physically emulated using transmission lines (parallel plate waveguides in our case) loaded with dielectric slabs, as shown in Fig. 1b. It is important to note that here we implement series junctions of T-circuits emulated by parallel plate waveguides as transmission lines. However, it is also possible to use Π-circuits emulated by transmission lines connected in a parallel[58–60] configuration, as it is explained in the Supplementary materials. The selection of T-circuits is to enable the implementation of the metatronic elements used to emulate the electronic circuits, while maintaining a small junction cross-section (compared to the operating wavelength of the structure), a requirement to ensure equal splitting of the incident signals at the junction according to Kirchhoff's laws[17,41,61]. Now, as the T-circuits from Fig. 1a (and their equivalent waveguide-based model from Fig. 1b) are connected to a junction in a series configuration, the flow of current through each of the connected T-circuits will form a *rotation* around the center of the junction (see Fig. 1a). Consider a junction in Fig. 1a where there is a *rotating* current $I_0$. By looking into one of the connected T-circuits at this junction, for example the top T-circuit, one can calculate the voltage $V_1$ across the circuit (at the input of the junction, see Fig. 1) by considering the difference between the *rotating* current at the central junction $I_0$ and the top junction $I_1$ as $V_1 = Z_p(I_1 - I_0) - Z_s I_0$, where $Z_p$ and $Z_s$ are the parallel and series impedance values of the T-circuit, as shown in Fig. 1a,b (full details of this calculation can be found in Supplementary materials). The magnitude and direction (clockwise/counterclockwise) of the *rotating* current at the junction can be found by solving Kirchoff's voltage law ($\sum_a V_a = 0$) at the junction using the associated voltages from each T-circuit, with $a$ representing the T-circuit from which the voltage is obtained ($a = 1, 2, 3, 4$, each representing the top, right, bottom and left T-circuit, respectively). After accounting for the division of current at the T-circuits, one can arrive at the following expression relating the rotating current at a junction to the rotating currents at each of the adjacent junctions in the network (see Supplementary Materials for the full calculation):

$$Z_p(I_1 + I_2 + I_3 + I_4 - 4I_0) - 4Z_s I_0 = 0 \qquad (4)$$

where $I_1, I_2, I_3, I_4$ and $I_0$ are the rotating currents at the top, right, bottom, left and center junctions, respectively. Now, it can be seen how Eq. 3 and Eq. 4 are analogous if the impedance values for the elements forming the T-circuit are chosen such that $Z_p = 1/h^2$ and $-4Z_s = k^2$. Hence, as the first term of Eq. 4 is analogous to the Laplacian, similar to Eq. 1 as explained above, while the second term is analogous to the zeroth-order term of the PDE, the proposed structure could indeed be used to calculate the solution of the Helmholtz equation. Importantly, while the addition of $Z_p$ and $Z_s$ to the network can enable us to establish an analogy between Eq. 3 and Eq. 4, they require a further transformation to be strictly valid as $Z_p$ in Eq. 4 is generally complex valued while $h$ in Eq. 3 is strictly real. This can be tackled through a transformation of the calculated currents such that $I'_a = I_a/Z_p^*$, where $Z_p^*$ is the complex conjugate of the parallel impedance and $I'_a$ is the transformed current value around the connected junction $a$. By substituting this into Eq. 4, the equation governing the transformed current distribution can be written as follows:

$$Z_p Z_p^* (I'_1 + I'_2 + I'_3 + I'_4 - 4I'_0) - 4Z_s Z_p^* I'_0 = 0 \qquad (5)$$

where $I'_1, I'_2, I'_3, I'_4$ and $I'_0$ are the transformed currents at the top, right, bottom, left and center junctions, respectively. Eq. 5 is analogous to Eq. 3 if $Z_p$ and $Z_s$ are now selected such that $Z_p Z_p^* = |Z_p|^2 = 1/h^2$ and $-4Z_s Z_p^* = k^2$, which are different from the impedances defined to emulate Eq. 3 using Eq. 4 due to the transformation of the currents described above.

**Emulating T-circuit lumped elements via metatronic circuit elements**

In the previous section, it has been described how periodic T-circuits formed by lumped circuit elements within a network may be used to calculate the solution of PDEs when arranged in a grid. To implement these lumped elements at high-frequencies (microwaves in our case), here we opted for the alternative of exploiting metatronic elements: thin dielectric, or metallic inserts, that may be engineered to emulate the performance of lumped circuit elements at different frequency ranges (from microwaves up to visible)[43-46,62]. To mimic the series and parallel impedances of the T-circuit from Fig. 1 (i.e., $Z_s$ and $Z_p$, respectively) we consider a metatronic circuit formed by three thin films (either dielectric or metallic) embedded within a host medium (vacuum in our case, $\varepsilon_r = 1, \mu_r = 1$), as shown in Fig. 1b. When considering a transverse electromagnetic signal (TEM, as it is in the case of the proposed parallel plate waveguides) impinging onto a single one of these dielectric or metallic slabs under normal incidence, the slab can be defined by a parallel impedance with a value[46]

$$Z_p = \frac{i}{\omega \varepsilon_0 \varepsilon_r(\omega) w} \qquad (6)$$

where $\omega$ is the angular frequency of the incident wave, $\varepsilon_r$ is the relative permittivity of the thin material emulating a metatronic element, $w$ is the thickness along the propagation axis and $i$ the imaginary number. In Eq. 6 and all other calculations the time convention $e^{-i\omega t}$ is used. From this expression, a slab with a positive real permittivity value $[Re(\varepsilon_r) > 0]$ (a dielectric slab) will behave as a capacitive lumped element. However, if a dispersive material with a negative real permittivity is implemented $[Re(\varepsilon_r) < 0]$ it will instead behave as an inductive lumped element. Based on this, to emulate the series impedance $Z_s$ from Fig. 1a we made use of a thin layer of a material embedded in between two $\lambda_0/4$ free-space regions. While the thin layer alone will be able to emulate a metatronic element in parallel (i.e., a parallel impedance $Z_p$), placing it in between the two $\lambda_0/4$ free-space regions enables us to apply an impedance transformation[63] such that the isolated parallel impedance $Z_p$ of the metatronic element is transformed into an effective series impedance $Z_s = Z_0^2/Z_p$, (where $Z_0$ is the impedance of free-space $Z_0 = \sqrt{\mu_0/\varepsilon_0}$). By combining this expression of $Z_s$ with Eq. 6, one can arrive to the final value for the effective series impedance, as follows[46].

$$Z_s = -i\omega\varepsilon_0\varepsilon_r w Z_0^2 \qquad (7)$$

From Eq. 7, it can be seen how, due to the impedance transformation, materials with $Re(\varepsilon_r) > 0$ and $Re(\varepsilon_r) < 0$ are now expressed as series inductors and capacitors, respectively. Based on this, we can emulate the required T-circuit to solve the PDE described in the previous section (see Fig. 1a and Eq. 5) by designing the metatronic elements using Eq. 6 and Eq. 7. The proposed metatronic element-based structure that can emulate a lumped-element based T-circuit is shown in Fig. 1b. The metatronic circuit consists of a parallel plate waveguide having two thin slabs (yellow slabs) placed in between of two $\lambda_0/4$ regions filled with air, in so doing the series impedances of the T-circuit are emulated. For the parallel impedance, a thin slab is also used (brown slab) as shown in Fig. 1b. This structure is then arranged in a network configuration forming junctions as explained above (schematically shown in Fig. 1a). Importantly, as the dielectric slabs are embedded within parallel plate waveguides, (see Fig. 1b) Eq. 6 and Eq. 7 for the calculation of metatronic elements are valid in our case as we are also working with an incident transverse electromagnetic TEM signal. Now, when selecting the dimensions of the waveguides in which the dielectric slabs are embedded, one must ensure that the separation between plates is small compared to the incident wavelength in order to limit the impact of fringing fields at the edges of the waveguide (small cross-sections are also required when working with waveguides connected in a parallel configuration[58,59,64], as it is the case shown in the Supplementary Materials). Finally, in this work we consider that all waveguides have the same filling materials (vacuum $\varepsilon_r = 1$, $\mu_r = 1$) and cross section (i.e., equal impedance).

With this configuration, the numerical results of the out-of-plane $H_z$-field distribution, calculated at the junctions, for a $3 \times 3$ waveguide network are shown in Fig. 1c. A monochromatic (10 GHz, $\lambda_0 = 30$ mm) incident signal is excited from the left waveguide of the top-left junction, here labeled as junction 1 in Fig. 1c. Using Eq. 6 and Eq. 7, the metatronic elements were designed to emulate the parallel and series impedance values of $Z_p = 2.5iZ_0$ and $Z_s = -0.9iZ_0$, respectively. These impedances correspond to $h$ and $k$ values of 0.4 and 3 (arbitrary units), respectively. It should be noted that the units of $h$ and $k$ are informed by the physics of the equation being solved. For example, in an EM problem, $h$ and $k$ have units of meters (m) and m$^{-1}$, respectively. However, here we first consider both of them as unitless in order to calculate a general solution of a PDE. With these considerations, the metatronic elements have a thickness of 0.2 mm ($\lambda_0/150$) each, with permittivity values of $\varepsilon_r = 9.54$ and $\varepsilon_r = 21.44$ for the slabs emulating parallel and series impedances, respectively. To corroborate the T-circuit design, a full ABCD matrix analysis[3] of the structure presented in Fig. 1b (upper panel) was performed (see models in Supplementary Materials). Based on these calculations the reflection and transmission coefficients varied slightly from the ideal design using the T-circuit model, as expected due to the non-zero thickness of the dielectric slabs for the structure in Fig. 1b. To overcome this difference, one can simply optimize the geometrical or EM parameters of the structure shown in Fig. 1b by, for instance, slightly changing the distance $\lambda_0/4$ between the slabs, by changing $w$ of the slabs or by modifying $\varepsilon_r$ of the slabs. However, in this first example, the structure from Fig. 1b is kept unchanged with the same dimensions as described above, but now we can retrieve the equivalent $Z_p$ and $Z_s$ values from the ideal T-circuit model that match the reflection and transmission coefficients from the ABCD method of Fig. 1b (details on this calculation can be found in the Supplementary Materials). The calculated impedances are $Z_p = 2.1522 i Z_0$ and $Z_s = -0.9311 i Z_0$, respectively (corresponding to $h = 0.4646$, $k = 2.8313$, i.e. a small variation of the PDE parameters). In this example, these new values of $Z_p$ and $Z_s$ are then used to calculate the theoretical values for the $H_z$-field based on the ideal circuit model and these results are compared to the numerically calculated results using simulations, as it will be shown below.

The numerical $H_z$-field values (amplitude and phase) at junctions 1 through 9, are presented in the middle panel of Fig. 1c. These simulations were obtained using the full-wave simulation software CST Studio Suite® (see the methods section for details of the simulation setup). As shown in Fig. 1a and discussed above, the solution of a PDE can be obtained by looking at the $H_z$-field (and hence the circulating current) at the center of a waveguide junction. However, as Eq. 2 requires four terms from neighboring junctions [$g(x, y + h)$, $g(x + h, y)$, $g(x, y - h)$ and $g(x - h, y)$ from the top, right, bottom and left junctions, respectively], only a junction which is fully surrounded by four other junctions will be able to produce a solution to the PDE. By looking at the structure from Fig. 1c (left panel) only junction 5 of the $3 \times 3$ network will satisfy Eq. 5. The phasor representation of the numerically calculated values for the $H_z$-field at junction 5 and its neighboring junctions are presented in the rightmost panel of Fig. 1c as "×" symbols, alongside the theoretical values for an ideal grid of metatronic elements calculated using the T-circuit model with the impedance values obtained via the ABCD matrix method (i.e., $Z_p = 2.1522iZ_0$ and $Z_s = -0.9311iZ_0$ ), represented as "○" symbols (see supplementary materials, as mentioned above). As observed, an excellent agreement between the numerical and theoretical results is obtained. Finally, the numerical results of the $H_z$-field for the adjacent junctions can also be used to calculate what would be the ideal value for the $H_z$-field at junction 5 that will satisfy Eq. 3. This can be calculated as $H_5 = (H_2 + H_4 + H_6 + H_8)/(4 - h^2k^2)$, where the subscripts indicate the junction number (see Fig. 1c). By doing this, the error between the numerical solution and the ideal solution of the Helmholtz equation can be calculated as the difference between the numerical results for $H_z$-field at junction 5 and the ideal value of $H_5$ that satisfies Eq. 3 (as described before). The difference between these results is ~7.23%, demonstrating how a small error is achieved and the proposed structure is capable of calculating a solution to the Helmholtz equation at junction 5.

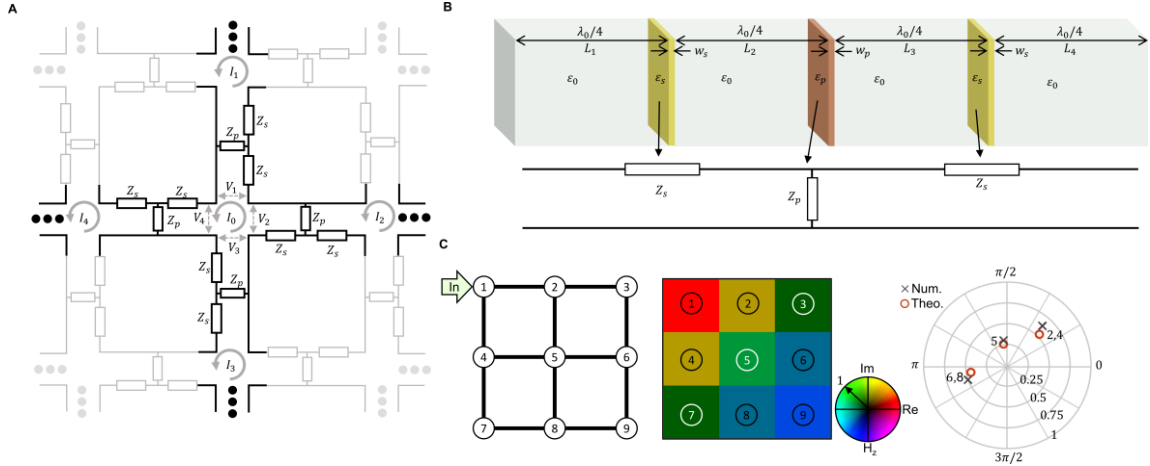

**Fig 1: Transmission line schematic representation of metatronic loaded network to solve PDEs. a,** Equivalent circuit representation of a single node of the proposed analogue processor. Each node is connected to four adjacent nodes via a T-circuit. In this representation the out-of-plane $H_z$-field at the center of each junction ($H_a$, where $a$ indicates the junction number) is represented by the current flowing around that junction counterclockwise ($I_a$). **b,** (top) waveguide-based metatronic structure that can emulate a T-circuit from **a**. It consists of three thin dielectric slabs separated by a distance $\lambda_0/4$. (bottom) equivalent T-circuit model. **c,** Full-wave numerical simulation results of a $3 \times 3$ network of waveguide-based metatronic structures emulating T-circuits with $Z_p = 2.1522iZ_0$ , $Z_s = -0.9311iZ_0$ corresponding to $h = 0.4646$ and $k = 2.8313$. (left) Nodal representation of the simulation setup with a single monochromatic (10 GHz) incident signal excited from the top left waveguide junction (labeled 1). (middle) Simulation results of the out-of-plane $H_z$-field (amplitude and phase) extracted at the center of each waveguide junction (labeled as 1-9). These results are normalized such that the input signal at junction 1 is unity. (right) numerical (black crosses) and theoretical (red hollow circles) results of the phasor values of $H_z$-field recorded at junctions 2,4,5,6, and 8.

**Scaling of the calculated PDE solutions.**

As discussed in Fig. 1c, the proposed structure can be used to calculate the solution of a PDE at the junction of waveguides as long as the junction is interconnected to adjacent junction in all directions (for instance junction 5 in Fig. 1c). Hence, if one wants to calculate a more detailed solution to the PDE (Helmholtz equation in this manuscript, for instance), it is necessary to increase the size of the network by introducing more junctions that are interconnected by waveguides filled with slabs (to emulate T-circuits) at their top, bottom, left and right directions. In other words, if a sampling point of the solution of the PDE is represented by a junction, it is possible to increase the resolution of the calculated solution by increasing the number of junctions. To address this, here we consider networks of waveguide-based metatronic T-circuits consisting of $N \times M$ junctions along the horizonal and vertical directions, respectively (see Fig. 2a). A three-dimensional (3D) representation (top view) of a central section of a larger network of junctions is included in Fig. 2b, showing how there are now more junctions fully interconnected to adjacent junctions in all directions compared to the scenario discussed in Fig. 1c. As explained before from Eq. 5, the scaling and spatial sampling of the calculated solution is determined by the chosen $h$ value of the structure. This means that it is possible to obtain the solution of a PDE with an increased resolution by increasing the physical size of the network (number of junctions) while keeping the same simulation space. This will effectively represent a smaller separation between sampling points in the simulation space $h$. Alternatively, a higher resolution of the solution for a PDE could be achieved without altering the size of the network but instead reducing $h$. This will mean that the simulation space will become smaller, as it will be discussed below.

Consider for example, an $N = M = 25$ network of junctions where 96 out of the total 625 junctions are placed at the edges and used as boundary junctions; i.e., as they are not surrounded by junctions in all directions, they cannot be used as part of the solution of the PDE. The analytical (calculated using the Huygens-Fresnel principle), theoretical (calculated by solving the governing equations of an ideal T-circuit-based network) and numerical (CST Studio Suite®) results of the out-of-plane $H_z$-field distribution at the junction of waveguides emulating the T-circuits are shown in Fig. 2d. Here we consider the incident signal as in Fig. 1c (10 GHz monochromatic incident signal excited at the left waveguide of the top left junction). The metatronic elements were designed using Eq. 6 and Eq. 7 to emulate the parallel and series impedance values $Z_p = 2.5iZ_0$ and $Z_s = -0.9iZ_0$ (impedances corresponding to $h = 0.4$, $k = 3$). After optimizing the geometric (distance between slabs, $w_s$ and $w_p$) and EM parameters (permittivity values of the slabs $\varepsilon_r$) of the design, the emulated impedance values calculated from the ABCD matrix method where $Z_p = 2.498iZ_0$ and $Z_s = -0.9003iZ_0$ ($h = 0.4003$, $k = 2.999$), respectively (see dimensions and EM parameters in the caption of Fig. 2). As it is shown in Fig. 2d, there is an excellent agreement between all the results with the solution of the PDE resembling the radiation of a dipole. Note that this is an expected result due to a monochromatic signal is applied only from the top-left junction. As a result, the analytical calculations using the Huygens-Fresnel principle will only have a single point as the radiating source, as observed in Fig. 2d.

For completeness, and in order to demonstrate the impact of the chosen $h$ value as a scaling parameter, the metatronic elements were also modified so that the calculated solution of the PDE would resemble a zoom-in image of the top-left quadrant of the solution calculated in Fig. 2d. To do this, the metatronic elements were designed to emulate the impedance values $Z_p = 5iZ_0$ and $Z_s = -0.45iZ_0$ ($h = 0.2$, $k = 3$). As observed, now $Z_p$ and $Z_s$ have been doubled and halved (compared to the values for Fig. 2d), respectively. This in turn can enable a solution to the same PDE as Fig. 2d (same $k$ value) but now with the separation between sampling points $h$ being halved. As a result, the simulation space is reduced to a quarter, effectively zooming into the top-left quadrant of Fig. 2d (dashed white square). The emulated impedance values after optimization (see dimensions, EM parameters in the caption of Fig. 2) are $Z_p = 5.001iZ_0$ and $Z_s = -0.4501iZ_0$ ($h \approx 0.2$, $k = 3.001$) and the results are shown in Fig. 2e. By observing the number of oscillations present in Fig. 2e and comparing it to the top-left quadrant of Fig. 2d, it can be qualitatively confirmed that the results presented in Fig. 2e indeed represent a zoomed-in solution to the same PDE as Fig. 2d. Furthermore, as the same number of sampling points (junctions) are used to display a smaller area in the simulation space (one quarter of Fig. 2d), the resolution of the calculated solution is quadrupled. For completeness, line plots showing the analytical, theoretical and numerical magnitude of the $H_z$-field along a line from the top-left to bottom-right of each simulation space are shown in the last column of Fig. 2d,e. In these figures all three plots show the characteristic 1/distance decay associated with a radiating dipole in 2D. However, in the numerical and theoretical plots there are oscillations. As will be discussed in more detail later, this is due to the impact of reflections produced at the boundary junctions. Finally, to demonstrate that the waveguides along with the loaded dielectric slabs are indeed emulating the T-circuits from Fig. 1a and are the responsible for the PDE solving performance of the proposed network, the case when the waveguides are not loaded with the dielectric slabs is also shown in Fig. 2c. This is done by replacing the dielectric slabs with air, so that the physical length of the connection between junctions is the same as in Fig. 2d,e. As observed the solution does not resemble the solutions to the Helmholtz equation shown in Fig. 2d,e indicating that it is indeed the dielectric slabs and the waveguides (as a whole) that are responsible for the PDE solving behavior of the structure.

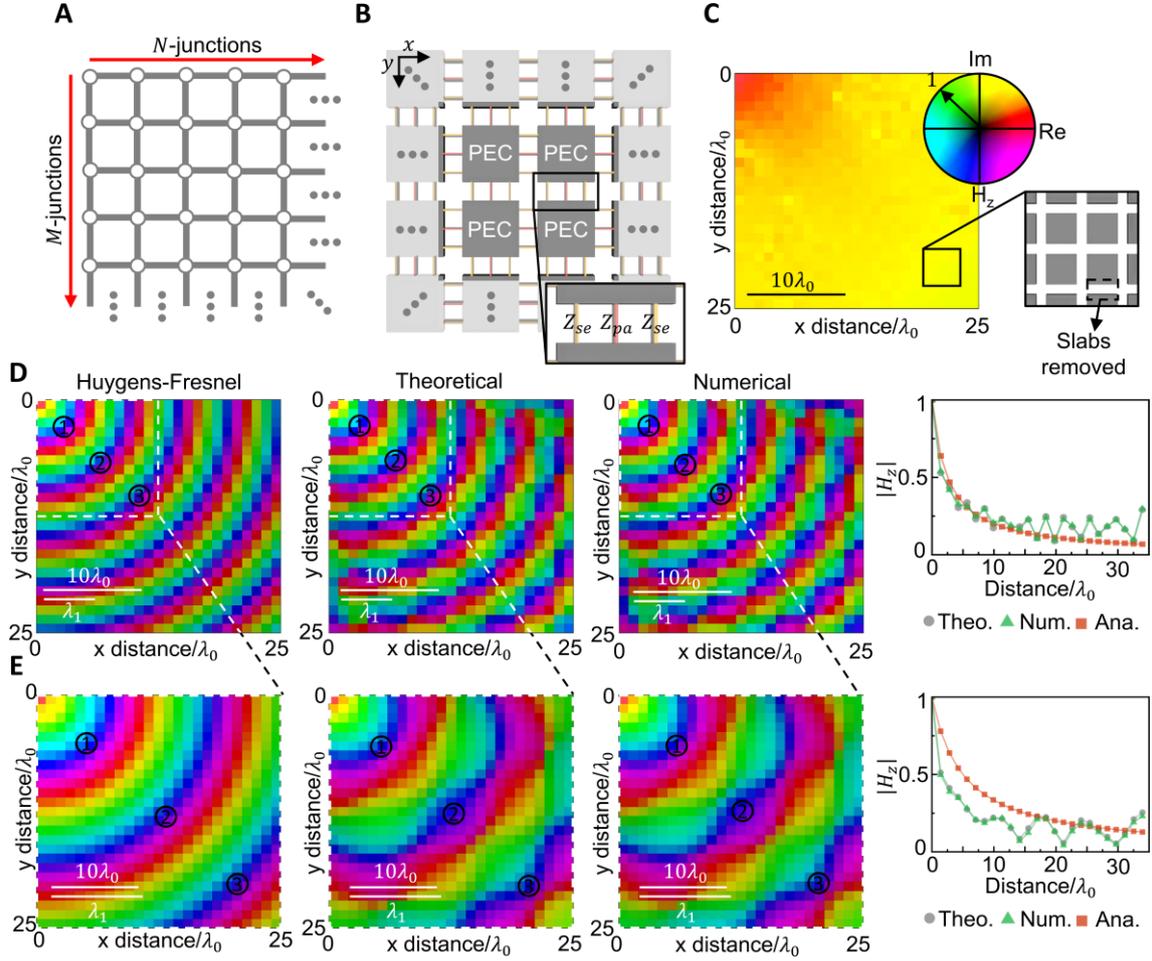

**Fig 2: Scaling of PDE results. a,** Nodal representation of an arbitrarily sized network of interconnected metatronic elements. **b,** (top view) 3 × 3 section of a larger waveguide network extending in all directions. The perfect electric conductor (PEC) regions for the waveguides are represented by as grey blocks. The red and yellow slabs represent the elements that enable us to emulate the parallel and series metatronic elements respectively. **c-e,** full-wave numerical results of the $H_z$-field distribution of a 25 × 25 waveguide network. As in Fig. 1c, a 10 GHz monochromatic input signal is excited from the left waveguide of the top-left junction. These results have been normalized so that the out-of-plane $H_z$-field seen at the top left junction is unity. Here, $\lambda_1$ is the wavelength of the PDE in the simulation space, while $\lambda_0$ is the wavelength of the incident signal in free-space. **c,** Numerical results for the case where there are no dielectric slabs present within the connecting waveguides, see inset. **d, e,** Analytical (left), theoretical (middle) and numerical (right) results of the same setup from panel **c** (same color scale applies here) but now when the waveguides are loaded with the dielectric slabs emulating metatronic elements with effective impedances $Z_p = 2.498iZ_0$, $Z_s = -0.9003iZ_0$ and $Z_p = 5.001iZ_0$, $Z_s = -0.4501iZ_0$, respectively. These values correspond to PDEs with parameters $h = 0.4003$, $k = 2.999$ ($\lambda_1 = 2.095$) and $h \approx 0.2$, $k = 3.001$ ($\lambda_1 = 2.094$) respectively. The EM and geometrical parameters after optimization are $L_1 = L_4 = 7.394$ mm (~$0.2465\lambda_0$), $L_2 = L_3 = 7.307$ mm (~$0.2436\lambda_0$), $w_s = 0.2111$ mm (~$7.0366\times10^{-3}\lambda_0$), $w_p = 0.1741$ mm (~$5.8033\times10^{-3}\lambda_0$), $\varepsilon_s = 21.5$, $\varepsilon_p = 12$ for the results presented in **d**, where $L_1, L_2, L_3$ and $L_4$ are the lengths of the impedance transforming waveguides from left to right, as shown in Fig. 1b. $w_s$, $w_p$, $\varepsilon_s$ and $\varepsilon_p$ are the widths and permittivity values of the slabs representing the series and parallel elements, respectively. These parameters are $L_1 = L_4 = 7.390$ mm (~$0.2463\lambda_0$), $L_2 = L_3 = 7.294$ mm (~$0.2431\lambda_0$), $w_s = 0.2201$ mm (~$7.3367\times10^{-3}\lambda_0$), $w_p = 0.1911$ mm (~$6.37\times10^{-3}\lambda_0$), $\varepsilon_s = 10.80$, $\varepsilon_p = 6.000$ for the results presented in **e**. The line plots in the rightmost panels of **d, e** show the numerical (green triangles), analytical (red squares) and theoretical (grey circles) results of the magnitude of the $H_z$-field taken along a straight line from the top-left to bottom-right corners of the simulation space, respectively.

**Calculating solutions to Dirichlet boundary value problem.**

The examples presented in the previous sections have considered only a single input signal applied from a waveguide connected at one boundary junction (top-left junction from Fig. 1 and Fig. 2). Here, it is shown how the proposed structure for PDE solving can also be used to calculate the solutions to Dirichlet boundary value problems[65]. This is done by implementing simultaneous excitations from the different outer waveguides connected to the boundary junctions (from now on called boundary waveguides) around the network. For this, the left waveguide from the top-left junction is used as a reference. In this way, Dirichlet boundary conditions, such as $g = 1$ or $g = 0$, can be considered at each of the boundary junctions. These boundary conditions are physically implemented in the structure from Fig. 1 by designing the inputs from the boundary waveguides such that the $H_z$-field at the center of each of the boundary junctions is 1 or 0, respectively. To implement a specific boundary condition at a boundary junction, it is important to carefully engineer the amplitude ratios and phase differences of the incident signals applied from the different boundary waveguides. This means that the boundary value at a given boundary junction (calculated by the $H_z$-field at the junction center) is determined by the superposition of the input signal applied from the connected boundary waveguide and those signals coming from the other boundary junctions from the network.

The required incident signals from the boundary waveguides can be directly calculated from the scattering matrix of the structure $\boldsymbol{A}$[3,58,60,64]. As an example, consider a vector of incident signals defined by their $E$-field (the same process could be done considering voltages) $\boldsymbol{x} = [x_1, x_2 \ldots, x_{2(M+N)}]^T$ where $T$ indicates the transpose operation. These signals are used to excite the boundary waveguides connected to the boundary junctions. Due to the interaction of these incident signals with the PDE solving structure, a vector of output signals defined by the $E$-fields $\boldsymbol{y} = \boldsymbol{A}\boldsymbol{x}$[58,60,64] is created (observed at the same boundary waveguides), with $\boldsymbol{y} = [y_1, y_2, \ldots, y_{2(N+M)}]^T$. The vector containing the complex values of the instantaneous $H_z$-field (which relates to the current rotating around the junction, as explained above) at each of these boundary junctions can be written as[3] $\boldsymbol{H}_b = (\boldsymbol{x} - \boldsymbol{y})/Z_0$, where $Z_0$ is the characteristic impedance of the boundary waveguides (as before, all the waveguides in the network have the same dimensions and filling materials, i.e. they have the same characteristic impedance) and $b = [1, 2, \ldots 2(N + M)]$ represents each of the boundary waveguides connected to the boundary junctions. Combining this expression for $\boldsymbol{H}_b$ with the scattering equation for $\boldsymbol{y}$ allows us to define a relation between the desired boundary values at the boundary junctions and the vector of incident signals from the boundary waveguides required to produce them, as follows:

$$x = Z_1(I - A)^{-1}H_b \qquad (8)$$

where $I$ is the identity matrix with size $2(N + M)$, i.e., the total number of input waveguides at the boundary junctions.

With this configuration, two examples of Dirichlet boundary value problems which can be solved by our proposed structure using EM waves are presented in Fig. 3. For both cases, we exploit the same $25 \times 25$ network of waveguide-based metatronic structures as in Fig. 2e (right panel). In the first example (Fig. 3a,b) the incident signals from the boundary waveguides have been selected using Eq. 8 to produce a Dirichlet boundary value of $g = 1$ at the left boundary, while a value of $g = 0$ is implemented for the top, right and bottom boundary junctions, represented by the red and greyscale lines in Fig. 3a respectively. These boundary values mean that the calculated solution to the PDE will resemble a standing wave produced when a wave propagates from left to right and is reflected by the right boundary. For the analytical calculations in Fig. 3, the solution of the PDEs are calculated using the FEM from the build in PDE toolbox of a commercial software[66,67] (MATLAB in our case, see the methods section for more details about this calculation). To further compare the results shown in Fig. 3b, the $H_z$-field values were extracted along the horizontal and vertical lines (at 13 junctions down from the top boundary and 12 junctions along from the left boundary, respectively) as shown in the top panels from Fig. 3b and the results are shown in the bottom-left and bottom-right panels respectively, demonstrating an agreement between the analytical, theoretical and numerical results. The second example of a Dirichlet boundary value problem is presented in Fig. 3c, in which the incident signals are selected such that the magnitude of the $H_z$-field at each of the boundary junctions is the same. However, the phase is varied such that, along a closed path around the entire boundary, the $H_z$-field completes a full $2\pi$ phase loop. The analytical, theoretical, and numerical results of the $H_z$-field are shown in the top panels from Fig. 3d. For completeness, the magnitude of the $H_z$-field was recorded along the diagonal lines from these panels and the results are shown on the bottom panels from Fig. 3d, again, showing an agreement between the results and demonstrating the potential of the proposed structure for PDE solving.

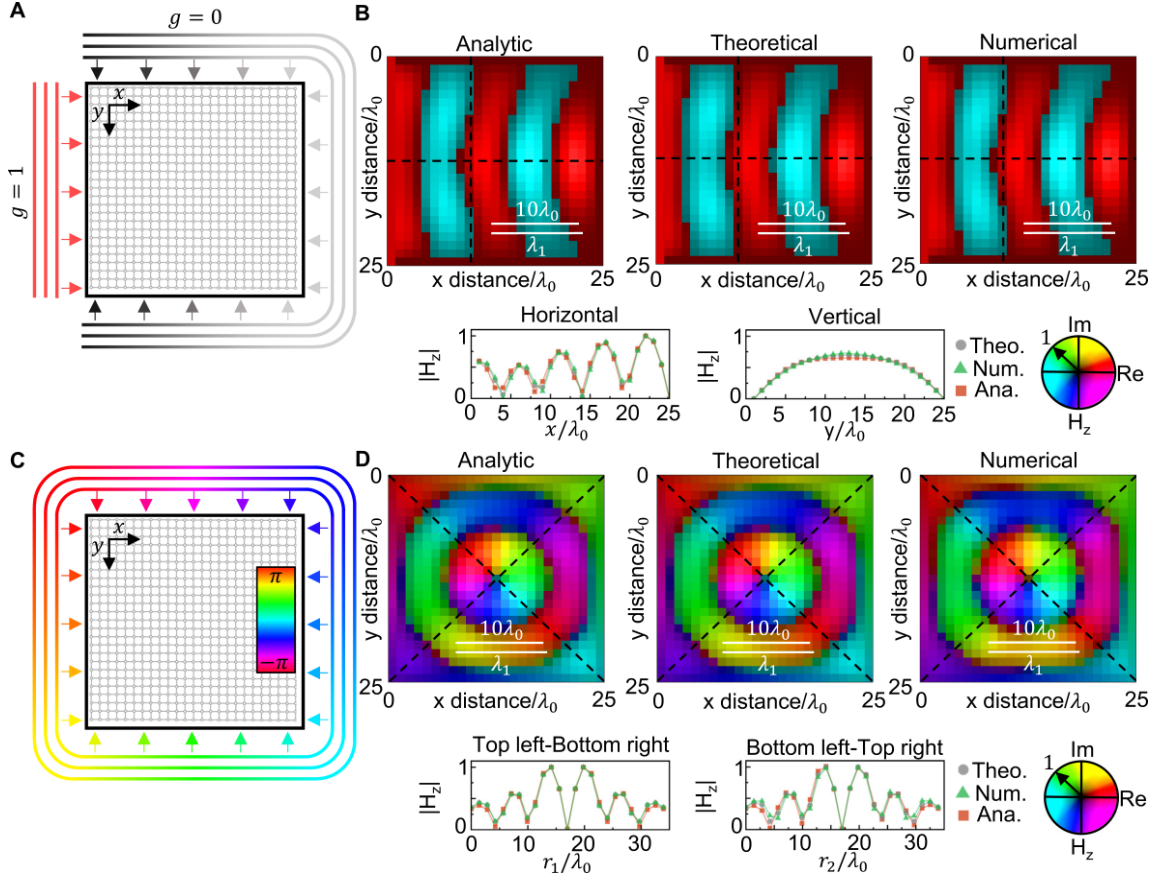

**Fig 3: Solving Dirichlet boundary value problems. a,c,** Schematic representations of a $25 \times 25$ network of junctions of waveguide-based metatronic circuits excited with different Dirichlet boundary conditions. The metatronic elements are chosen considering $Z_p = 5.001iZ_0$ and $Z_s = -0.4501iZ_0$ (see Fig. 2 caption for EM and geometrical parameters), corresponding to $h \approx 0.2$ and $k = 3.001$ ($\lambda_1 = 2.094$). **a,** The left-hand boundary is set to $g = 1$ while the top, right and bottom boundaries are $g = 0$. **c,** The boundary conditions are such that the magnitude at each boundary junction is 1 but the phase along the boundary spatially varies from 0 to $2\pi$, anticlockwise. **b,d,** Analytical, theoretical and numerical results of the scenarios from **a** and **c,** respectively. The top panels from **b,d** represent the normalized instantaneous $H_z$-field values calculated at each of the junctions inside the network. The bottom panels from **b,d** represent the magnitude of the calculated analytical (red squares), theoretical (grey circles) and numerical (green triangles) $H_z$-field values along the dashed lines from the top panels.

**Open boundary value problems.**

As a final study, here we show that it is also possible to use the proposed technique to solve open boundary value problems such as the two examples presented in Fig. 4. As it could be expected, the signals will need to be absorbed by the boundary waveguides mimicking an open boundary, similar to the well-known perfectly matching layers, PMLs[68]. In attempting to do this using the proposed metatronic-based network, by using waveguide ports connected to the boundary waveguides, unwanted large reflections can be obtained in the calculated solution. This is due to the impedance mismatch between the boundary junctions and the rest of junctions given that they are connected to 3 and 4 adjacent junctions, respectively. To tackle this, a solution is to extend the simulation space by including more junctions into the waveguide network and then extracting the PDE solution from a smaller portion of the simulation space, rather than the entire area. This allows for the signals within the network to decay as they propagate through the extended network, reducing the impact of reflections on the overall solution. With this configuration, we provide in Fig. 4 examples of open boundary value problems. Here we use a $150 \times 100$ metatronic network (with the same $Z_p$ and $Z_s$ values as the examples presented in Fig. 2e and Fig. 3, $Z_p = 5.001iZ_0$, $Z_s = -0.4501iZ_0$, $h = 0.2$ and $k = 3$). The simulation space used to produce the results in Fig. 4 is a $50 \times 50$ subnetwork of the total $150 \times 100$ network. The position of the subnetwork is selected such that the network extends for 50 junctions from the top, right and bottom boundaries of the subnetwork, respectively, effectively emulating the behavior of a PML at these boundaries. No extra junctions are connected to the left-hand boundary of the subnetwork, allowing for incident signals to be excited at the boundary waveguides connected to these junctions.

The theoretical and analytical results of the normalized power distribution calculated at the center of the junctions for the two scenarios from Fig. 4a,d are shown in Fig. 4b,e, respectively (numerical simulations are not shown due to the large size of the whole structure). The analytical results from Fig. 4b,e are calculated using the Huygens-Fresnel principle, considering each boundary junction as a radiating dipole and using the FEM from the PDE toolbox of a commercial software (MATLAB[66,67], as in Fig. 3), respectively. Let us first focus on the results from Fig. 4a-c which shows an example of focusing/lensing. In this case, the boundary waveguides connected to the boundary junctions on the left-hand side of the network are excited such that the $H_z$-field distribution along the boundary junctions resembles an output signal traveling away from a lens (designed to produce a focus at a position $x = 1.432\lambda_1$, $y = 2.387\lambda_1$, where $\lambda_1 = 2.094$ is the wavelength of the PDE. In the simulation space, the focus would appear 15 and 25 junctions along the $x$ and $y$ axes of the network, respectively, see Fig. 4a,b). As shown in Fig. 4b, a clear focus is produced at the positions $x = 1.623\lambda_1$, $y = 2.387\lambda_1$ (17 and 25 junctions along in $x$ and $y$, respectively), and $x = 1.432\lambda_1$, $y = 2.387\lambda_1$ for the theoretical and analytical calculations, respectively, demonstrating a good agreement. Note that from the theoretical results there are some slight spatial ripples. This can be attributed to the reflections produced at the boundary junctions of the entire network as mentioned above. As one would expect, the signal inside the network would asymptotically decay to zero if the network was infinitely long. However, as the network is finite (extended by 50 junctions from the top, right and bottom of the $50 \times 50$ subnetwork), reflections, while small, are still present. For completeness, the theoretical and analytical results extracted along horizontal and vertical lines from Fig. 4b are shown in Fig. 4c. As it is shown, despite the presence of small reflections in the PDE solution, the focus is still well defined with both results in good agreement.

A second and final example of an open value problem in shown in Fig. 4d-f where the simulation space has been divided in two regions. Here, an insert is considered to be placed within a background medium (with the latter being the same as the one used for Fig. 4a-c). The insert is modelled by removing the waveguide-based metatronic T-circuits (i.e. effectively removing any junctions within that region). By doing this, any incident signal upon this region will be reflected, enabling the junctions at the boundary to that region to act as boundary junctions with $g = 0$. Physically, this could be implemented by using a perfect magnetic conductor (PMC) such as doped epsilon near zero materials[69], or by replacing the metatronic T-circuits connecting to this region with PEC ended stubs of length $\lambda_0/4$. The insert, shown in Fig. 4d,e, is a $0.9549\lambda_1 \times 0.9549\lambda_1$ square centered at the middle of the simulation space, i.e., a $10 \times 10$ grid in the center of the $50 \times 50$ subnetwork. The theoretical and analytical results of the power distribution for this scenario are shown in Fig. 4e when considering a planewave excitation (boundary waveguides are excited such that the boundary junctions on the left-hand side of the network fulfill a boundary condition of $g = 1$). As observed, there is a clear agreement between the results demonstrating how the incident wave is reflected and scattered by the insert. For completeness the power distribution along the vertical and horizontal lines from Fig. 4e are shown in Fig. 4f. The small differences between the analytical and theoretical results are due to the finer mesh used in the FEM method for the analytical results compared to the number of junctions to discretize the simulation region using the network of waveguides. This is an expected result that may be reduced by increasing the resolution of the calculated PDE solution in the metatronic network using the methods discussed in the previous sections. However, the results from Fig. 4e,f can be considered as a good approximation of the analytical solution. The proposed technique can be translated to different frequency ranges such as the optical regime via the emulation of metatronic elements within waveguides with the ability to split the incident signal in all directions[42], allowing its potential experimental implementation in integrated photonic circuits.

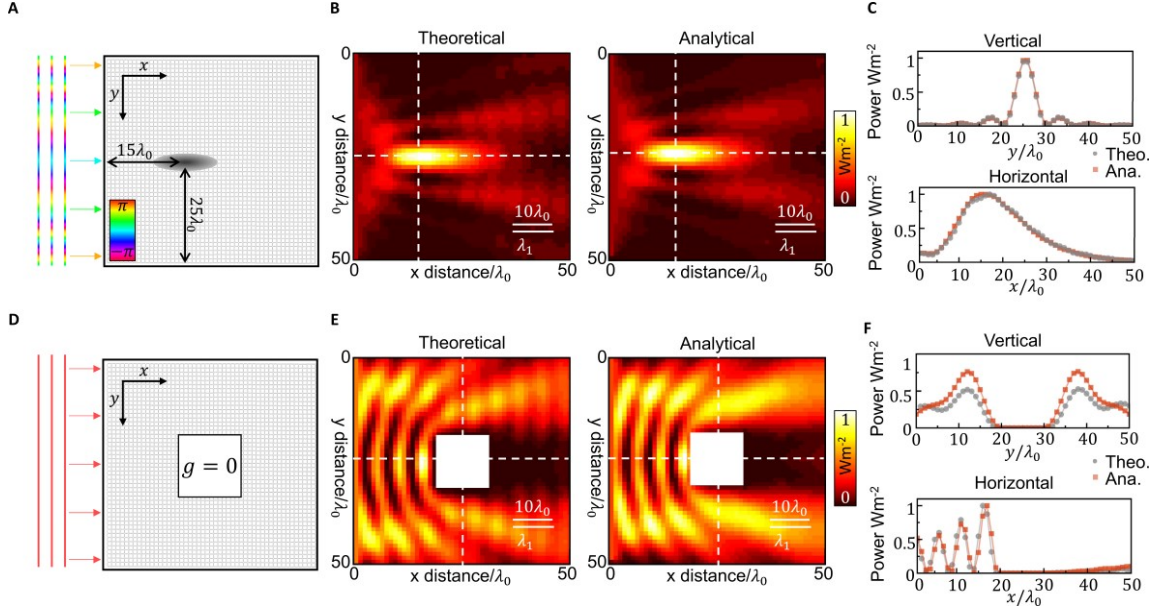

**Fig 4: Lensing and particle scattering. a,d** Schematic representations of a 50 × 50 subnetwork of waveguide-based metatronic circuits used to solve open boundary value problems. The 50 × 50 grid is constructed with $Z_p = 5.001iZ_0$ and $Z_s = -0.4501iZ_0$ corresponding to $h \approx 0.2$ and $k = 3.001$ ($\lambda_1 = 2.094$, where $\lambda_1$ is the wavelength seen inside the simulation space, see Fig. 2 caption for EM and geometrical parameters). **a**, A 10 GHz monochromatic wave is excited at each of the boundary waveguides connecting to the left-hand boundary junctions of the subnetwork. The amplitude and phase of these signals is selected such that the $H_z$ values at the boundary junctions resembles the output signal from a lens designed to produce a focus at $x = 1.432\lambda_1$, $y = 2.387\lambda_1$ (15 and 25 junctions in the $x$ and $y$ directions, respectively), represented by a grey spot. **d**, As in **a**, a 10 GHz monochromatic signal is excited at the left-hand boundary waveguides, now with amplitude and phase selected such that the $H_z$ values at the boundary junctions resembles a planewave. A $0.9549 \times 0.9549\lambda_1$ (10 × 10 junctions) $g = 0$ insert is placed at the center of the simulation space, represented by a white block. **b** and **e**, Theoretical (left) and analytical (right), power distribution for the scenarios presented in **a** and **b**, respectively. In **b** and **e**, results are normalized with respect to the power distribution at the focus and the maximum standing wave, respectively. **c,f**, Theoretical (grey circles) and analytical (red squares) values of power distribution along the vertical (top) and horizontal (bottom) lines drawn in **b** and **e**, respectively.

## Conclusions

In this work, an EM wave-based structure for analogue computing has been proposed. It consisted of a network of parallel plate waveguides loaded with carefully designed dielectric slabs such that the whole structure emulates a network of interconnected T-circuits, i.e., interconnected metatronic elements. It has been shown how the proposed structure has the ability to calculate the solution of PDEs in the form of the Helmholtz equation. Different scenarios have been demonstrated such as the calculation of Dirichlet (for example, wave propagation within a cavity) and open boundary value problems (including the solution for a focusing lens). The theoretical results have been compared with both analytical and numerical results demonstrating good agreement between them. The proposed technique can be translated to other spectral regimes by implementing specific waveguide-based structures for the desired frequency range (such as dielectric slab waveguides at optical frequencies) and the results here presented could open new opportunities for high-speed analogue computing and processors with light.

## Methods

The numerical results presented in Fig. 1-3 were obtained using the frequency domain solver of the commercial software CST Studio Suite®. Vacuum ($\varepsilon_r = 1$, $\mu_r = 1$) was used as the filling material of the waveguides from the network. The waveguides had a width of 0.1 mm ($\lambda_0/300$) and their length (separation between waveguide junctions) was determined by the parameters of the metatronic T-circuit as discussed in the main text above. Boundary conditions were set as *open* along the $x$ and $y$ axes (at the top, bottom, left and right boundaries), while the $z$ boundaries were both set to *magnetic* ($H_t = 0$). The metatronic T-circuits were implemented using three thin dielectric slabs immersed within the waveguides. Waveguide ports were placed at the ends of the boundary waveguides to excite the structure. The cross-section and filling materials of these boundary waveguides were the same as the waveguides described above. The boundary waveguides had a length of 3.75 mm ($\lambda_0/8$) between the ports and the boundary junctions. The out-of-plane magnetic field $H_z$ at the center of the waveguide junctions is extracted from the simulation using $H$-field probes placed at the center of each junction. The analytical results shown in Fig. 3,4e were obtained using the built in PDE toolbox from MATLAB by defining an elliptical PDE of the form $-\nabla[c\,\nabla g(x,y)] + a\,g(x,y) = f$, with $c = -1$, $a = k^2$ and $f = 0$. The simulation space in all cases was a square region with size ($Nh \times Mh$) with $h$ and $k$ as defined in the main text.

## Acknowledgements

V.P-P. and A.Y. would like to thank the support of the Leverhulme Trust under the Leverhulme Trust Research Project Grant scheme (RPG-2020-316). V.P-P. and R.G.M would like to thank the support from the Engineering and Physical Sciences Research Council (EPSRC) under the scheme EPSRC DTP PhD scheme (EP/T517914/1). For the purpose of Open Access, the authors have applied a CC BY public copyright license to any Author Accepted Manuscript (AAM) version arising from this submission.

## Conflicts of interests

The authors declare no conflicts of interests.

## Data availability

The datasets generated and analyzed during the current study are available from the corresponding author.